\documentclass[letterpaper, 10 pt, conference]{ieeeconf}  % Comment this line out
\usepackage{hyperref}
\IEEEoverridecommandlockouts                              % This command is only
\usepackage[utf8]{inputenc}
\usepackage[T1]{fontenc}
\usepackage{graphicx}
\graphicspath{ {./figures/} }

\title{\LARGE \bf
Input Agnostic Deep Learning for Alzheimer's Disease Classification Using Multimodal MRI Images}

\author{Aidana Massalimova and Huseyin Atakan Varol, \textit{Senior Member, IEEE}%
\thanks{A. Massalimova and H.A. Varol are with the Institute of Smart Systems and Artificial Intelligence, Nazarbayev University, 53 Kabanbay Batyr Ave., 010000, Nur-Sultan City, Kazakhstan. 
Email: \{aidana.massalimova, ahvarol\}@nu.edu.kz}
 \thanks{Corresponding author: Huseyin Atakan Varol.}}

\begin{document}

\maketitle
\thispagestyle{empty}
\pagestyle{empty}

%%%%%%%%%%%%%%%%%%%%%%%%%%%%%%%%%%%%%%%%%%%%%%%%%%%%%%%%%%%%%%%%%%%%%%%%%%%%%%%%
\begin{abstract}
Alzheimer's disease (AD) is a progressive brain disorder that causes memory and functional impairments. The advances in machine learning and publicly available medical datasets initiated multiple studies in AD diagnosis. In this work, we utilize a multi-modal deep learning approach in classifying normal cognition, mild cognitive impairment and AD classes on the basis of structural MRI and diffusion tensor imaging (DTI) scans from the OASIS-3 dataset. In addition to a conventional multi-modal network, we also present an input agnostic architecture that allows diagnosis with either sMRI or DTI scan, which distinguishes our method from previous multi-modal machine learning-based methods. The results show that the input agnostic model achieves 0.96 accuracy when both structural MRI and DTI scans are provided as inputs.   
\end{abstract}

%%%%%%%%%%%%%%%%%%%%%%%%%%%%%%%%%%%%%%%%%%%%%%%%%%%%%%%%%%%%%%%%%%%%%%%%%%%%%%%%
\section{Introduction}

Alzheimer's disease (AD) is a progressive neurodegenerative disease that causes age-dependent cognitive and functional decline. AD is the largest cause of dementia and accounts for 67\% of all cases. The number of patients diagnosed with AD was 46 million worldwide in 2019 and this number is forecast to reach 131 million by 2050. In the United States, AD is reported to be the sixth leading cause of death~\cite{alzheimer20192019}. Healthcare costs of AD patients are becoming a heavy burden for the already stretched healthcare budgets of the countries. Thus, it is vital to diagnose AD in the earlier stages to curb the associated healthcare costs and improve the quality of life of patients.

During the early stage of AD, the patients suffer from mild cognitive impairment~(MCI). Although a decline in cognitive function is characteristic of MCI, patients are capable to act independently in the activities of daily living~\cite{Hane2017}. Therefore, MCI is often mistaken with the symptoms of normal aging and is hard to detect in a clinical setting. It was reported that 44\% of individuals diagnosed with MCI later progressed to AD~\cite{alzheimer20192019}. 

The main imaging techniques used in studying AD are magnetic resonance imaging~(MRI) and positron emission tomography~(PET). Since the formation of neurofibrillary tangles (NFTs) and the deposition of extracellular amyloid-$\beta$ ($A\beta$) are major pathological hallmarks of AD, PET is used to detect and track these pathologies using various biomarkers. On the other hand, MRI allows investigating the structural connectivity changes~\cite{SoriaLopez2019}. Walhovd et al.~\cite{Walhovd2010} showed that measuring the hippocampal region using MR morphometry is more sensitive in diagnosing AD compared to PET imaging. 

\begin{figure}[b!]
\centering
\includegraphics[width=8cm]{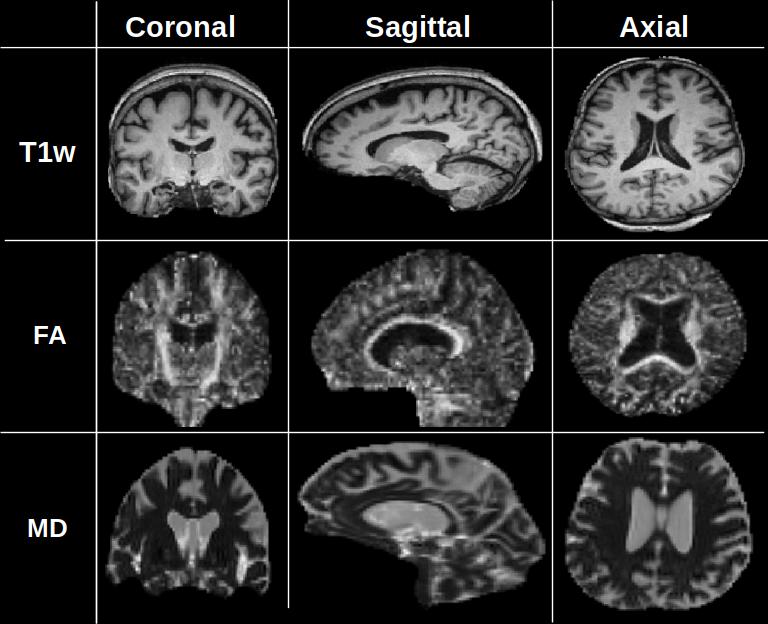}
\caption{Coronal, sagittal, and axial projections of skull-stripped T1w, FA and MD images from the OASIS-3 dataset. }
\label{figures:image_samples}
\end{figure}

Although AD is known more as a gray matter disease, recent studies revealed the white matter changes are prevalent to AD as well. White matter changes are conventionally investigated using diffusion tensor imaging (DTI) technique of MRI. DTI measures the diffusion of the water molecules along the fiber bundles. The diffusion direction is obtained from three eigenvalues ($\lambda_1$, $\lambda_2$, $\lambda_3$) and eigenvectors representing the water's molecular motion in three dimensions. The most common quantitative measures of DTI are fractional anisotropy (FA) and mean diffusivity (MD) (see Fig.~\ref{figures:image_samples}). FA measures the standard deviation of eigenvalues and is normally regarded as the general indicator of microstructural changes. On the other hand, MD is the average of directionality, inversely proportional to the membrane density and fluid viscosity \cite{alexander2007diffusion}. Decreased FA and increased MD were reported in AD patients with respect to healthy controls~\cite{Sexton2011} . 

Along with neuroimaging, clinical neuropsychological examinations such as the Mini-Mental State Examination (MMSE) and Clinical Dementia Rating (CDR) are commonly used to determine the staging of AD as clinical auxiliary diagnosis methods~\cite{perneczky2006mapping}.

Thanks to recent advances in deep learning and open access medical datasets, convolutional neural networks (CNNs) are extensively used in AD diagnosis based on the neuroimaging data. With their multilayered topology, CNNs can automatically extract low-to-high level latent features of the disease pathology. There are various works on AD diagnosis based on both a single imaging modality \cite{Basaia2019, ding2019deep,li2019deep} and multi-modality \cite{aderghal2020improving}, \cite{lee2019predicting}, \cite{kang2020identifying}. The majority of these works used Alzheimer's Disease Neuroimaging Initiative (ADNI) dataset~\cite{mueller2005ways}. % However, the differences in their image preprocessing methods, deep learning methods, and evaluation metrics perplexes the comparative study. 

In this work, we present a multi-modal neural network based on the combination of pretrained residual networks on T1-weighted (T1w), FA and MD scans (both obtained from the DTI images). As the data source, we used the new Open Access Series of Imaging Studies (OASIS) dataset~\cite{lamontagne2019oasis}. The novelty of our network is its capability for AD diagnosis with any of the FA, MD or T1w images as input. Instead of creating three different networks (one for each imaging modality), we created a single "input agnostic" network to benefit from transfer learning between different imaging domains. Even though we used two imaging modalities from the OASIS dataset, our network architecture enables the use of different datasets such as ADNI and also different imaging techniques such as PET. Similar to the ResNet trained on ImageNet dataset ~\cite{deng2009imagenet}, our network can also be utilized for transfer learning in solving other neuroimaging problems using deep learning.

The rest of the paper is organized as follows: In Section \ref{sec:method}, we describe the data characteristics, preprocessing steps of T1w and DTI scans and the network design. The results of our architecture are presented and discussed in Section~\ref{resultsanddiscussion}. We summarize the findings of this work in Section \ref{conclusions}.

\section{Methodology} 
\label{sec:method}

\subsection{Dataset}

The data used in this work was extracted from the OASIS-3 dataset~\cite{lamontagne2019oasis}. This dataset includes MRI and PET images from 1098 subjects that were collected at the Washington University Knight Alzheimer Disease Research Center over 15 years. We used only T1w and DTI scans for the normal cognition (NC), MCI, and AD classes. Tables \ref{T1w_information} and \ref{dti_information} provide the number of images, subjects and their demographics, MMSE and CDR scores for T1w and DTI modalities, respectively. Please note that not all subjects have both T1w and DTI scans.

\begin{table}[b!]
\caption{Clinical Information of Subjects with T1w Images}
\label{T1w_information}
\begin{center}
\resizebox{1\columnwidth}{!}{
\begin{tabular}{lccccc}
\hline
 \multicolumn{1}{l}{} & Subjects & Images & Age & MMSE & CDR \\ \hline
\multicolumn{1}{l}{NC}  &326&526&64.93$\pm$9.39&29.10$\pm$1.32&0.03$\pm$0.12 \\ \hline
\multicolumn{1}{l}{MCI} &17 &21 &68.08$\pm$8.62&28.57$\pm$1.60&0.31$\pm$0.37     \\ \hline
\multicolumn{1}{l}{AD}  &89 &100&72.04$\pm$6.23&25.95$\pm$3.93&0.64$\pm$0.48     \\ \hline
\end{tabular}}
\end{center}

\end{table}

\begin{table}[t!]
\caption{Clinical Information of Subjects with DTI Images.}
\label{dti_information}
\begin{center}
\resizebox{1\columnwidth}{!}{
\begin{tabular}{lccccc}
\hline
   \multicolumn{1}{l}{}   & Subjects & Images & Age & MMSE & CDR \\ \hline
\multicolumn{1}{l}{NC}  &381&623&65.81$\pm$8.97&29.04$\pm$1.40&0.02$\pm$0.11 \\ \hline
\multicolumn{1}{l}{MCI} &9&13 &66.71$\pm$5.56&28.54$\pm$1.51&0.35$\pm$0.32     \\ \hline
\multicolumn{1}{l}{AD}  &105 &122&71.69$\pm$6.01&26.35$\pm$3.69&0.59$\pm$0.48     \\ \hline
\end{tabular}}
\end{center}
\end{table}

\subsection{Preprocessing of T1w Images}

Prepocessing of anatomical T1w images were performed with SPM12 software (Wellcome Trust Centre for Neuroimaging, United Kingdom). All images were co-registered to the Montreal Neurological Institute (MNI) template to establish a uniform coordinate space. The underlying method behind SPM12 co-registration tool is based on mutual information maximization, which is commonly used in both uni- and multi-modal image registration. Consequently, each image had a size of 256$\times$256$\times$256 with a spatial resolution of 1mm$\times$1mm$\times$1mm.

\subsection{Preprocessing of DTI Images}
The preprocessing of DTI scans were done with the Dipy library developed for DTI analysis in Python. DTI scans were corrected for eddy current distortions and extracted from the skull. Afterward, the diffusion tensor parameters, FA and MD, were computed using (\ref{eq:FA}) and (\ref{equation:MD}). Finally, MD and FA images were affinely co-registered to their corresponding T1w images for the cases when both scans for the subject were available. The MD and FA images had a size of 96$\times$96$\times$96 with a spatial resolution of 2mm$\times$2mm$\times$2mm. 

\begin{equation}
FA=\sqrt{\frac{1}{2}}\frac{\sqrt{(\lambda_1-\lambda_2)^2+(\lambda_2-\lambda_3)^2+(\lambda_3-\lambda_1)^2}}{\sqrt{\lambda_1^2+\lambda_2^2+\lambda_3^2}}    
\label{eq:FA}
\end{equation}

\begin{equation}
MD=(\lambda_{1}+\lambda_{2}+\lambda_{3})/3
\label{equation:MD}
\end{equation}
\subsection{Data Split}
After preparing registered T1w, FA and MD scans, data were split subject-wise. However, not all subjects contained both DTI and and T1w scans. Therefore, we split the subjects into three groups: 1) subjects with only T1w scan, 2) subjects with only DTI scan, and 3) subjects with both T1w and DTI scans.

\begin{table}[b!]\centering
\caption{Number of Images per Class Before and After  Balancing.}
\label{Table:datasplit}
\begin{tabular}{llcccccc}
\hline
                            &     & \multicolumn{2}{c}{T1w Only}                              & \multicolumn{2}{c}{DTI Only}                              & \multicolumn{2}{c}{T1w+DTI}                            \\ \hline
                            &     & \multicolumn{1}{c}{train} & \multicolumn{1}{c}{test} & \multicolumn{1}{c}{train} & \multicolumn{1}{c}{test} & \multicolumn{1}{c}{train} & \multicolumn{1}{c}{test} \\ \hline
Before & NC  & 118                       & 26                       & 193                       & 48                       & 308                       & 74                       \\ \cline{2-8} 
                            & MCI & 10                        & 3                        & 4                         & 1                        & 7                         & 1                        \\ \cline{2-8} 
                            & AD  & 18                        & 7                        & 40                        & 7                        & 59                        & 16                       \\ \hline
After  & NC  & 118                       & 26                       & 193                       & 48                       & 308                       & 74                       \\ \cline{2-8} 
                            & MCI & 110                       & 27                       & 192                       & 47                       & 308                       & 70                       \\ \cline{2-8} 
                            & AD  & 126                       & 28                       & 200                       & 49                       & 295                       & 80                       \\ \hline
\end{tabular}
\end{table}

We divided the data into five folds for cross validation. In each fold, 80\% of the data was used for training the network and 20\% was set aside for testing. Due to the low number of samples, the median slices of the sagittal, axial, and coronal views from each 3D scan were concatenated to generate a single three-channel image. As it can be observed from Table \ref{Table:datasplit}, the classes were unbalanced. There were many samples for NC class but few number of samples for MCI and AD classes. Therefore, we generated additional images for the MCI and AD classes using neighboring slices of the median ones for each view. Since ResNet18 has input size of 224$\times$224$\times$3 \cite{resnet}, we resized FA and MD images using nearest neighbor interpolation. In the training process, we also augmented the data by applying random horizontal and vertical flips.

\subsection{Network Architecture}
\begin{figure}[b!]
\centering
\includegraphics[width=1.0\columnwidth]{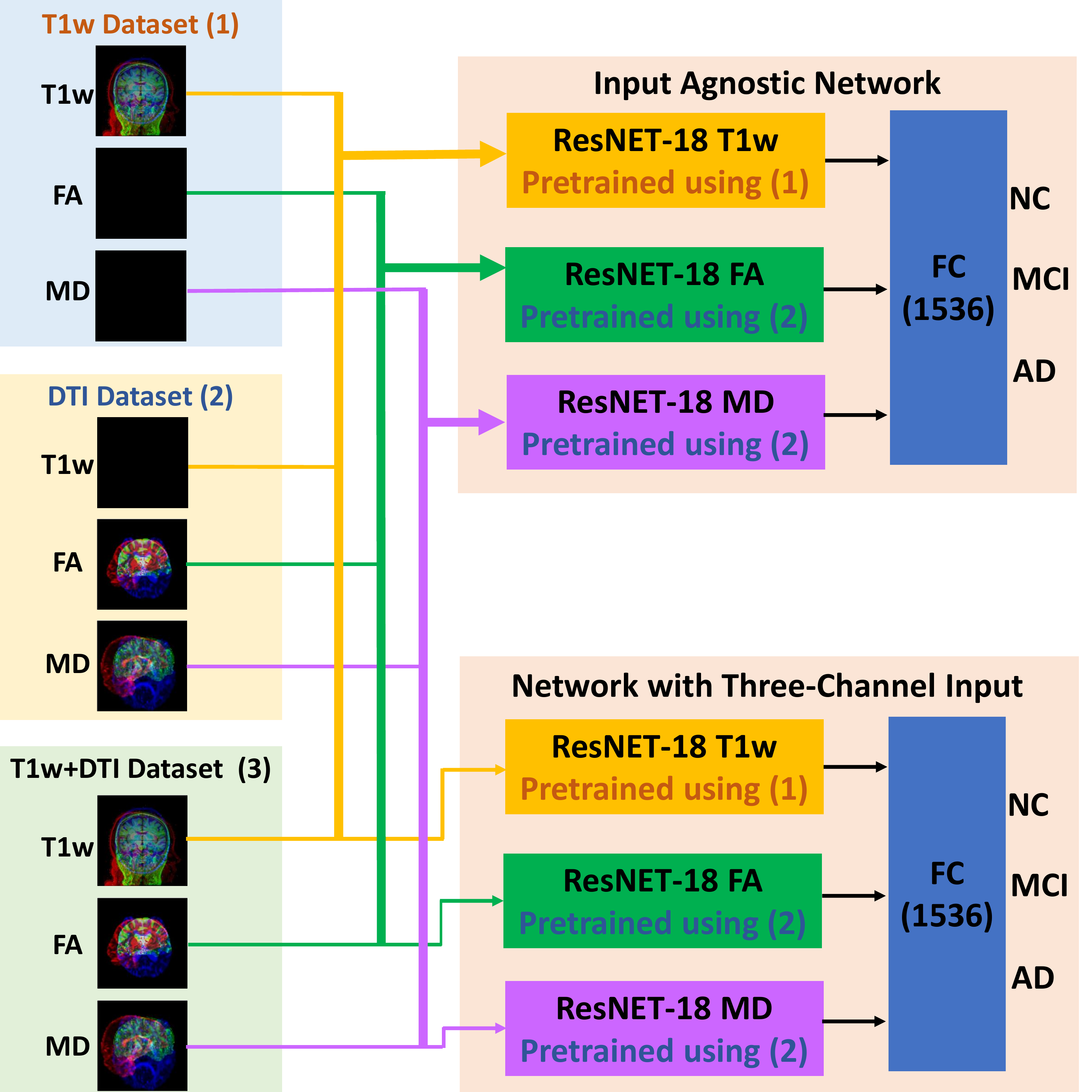}
\caption{Two network architectures for Alzheimer's disease classification: Input agnostic network capable and the network with three-channel (T1w, FA, and MF) input.}
\label{figures:architect}
\end{figure}

It was reported that transfer learning utilizing general image classification networks such as ResNet, LeNet and AlexNet outperforms non-transfer learning methods in AD classification \cite{aderghal2020improving}. In this work, we first trained three networks (for T1w, FA, and MD scans) using the pretrained ResNet18 as the base to classify NC, MCI, and AD classes. After fine-tuning each network, three models with best performance were selected to create the multi-modal network shown in the lower right part of the Fig.~\ref{figures:architect}. Since we need these three pretrained models for feature extraction from each input type, their fully-connected layers were removed and the weights of the remaining parts were frozen. The outputs of the last average pooling layer were fed to one common fully-connected layer with 1536 elements, which in turn classifies the images into the NC, MCI or AD classes. This model was trained on the data that included both T1w and DTI (FA and MD) scans, i.e. the third column in Table~\ref{Table:datasplit}. In this case, the network required all three image types as input. 

The second case aimed to enable the diagnosis using both (T1w and DTI) or either data types (T1w or DTI). We call this input agnostic network architecture (see upper right part of the Fig.~\ref{figures:architect}). We trained the input agnostic network using three different sets of data. The first dataset used T1W images from the T1w Only and T1w+DTI datasets (first and third columns of Table~\ref{Table:datasplit}). Similarly, the second dataset used FA and MD images from the DTI Only and T1w+DTI datasets (second and third columns Table~\ref{Table:datasplit}). Since the architecture requires three inputs, the other input images for cases containing only either T1w or DTI scans were substituted with black images. The third set was the T1w+DTI dataset (third column of Table~\ref{Table:datasplit}). 

For model training, the batch size was set to 16 and Adam optimizer was used~\cite{zhang2018improved}. The experiments were conducted on NVIDIA DGX-2 server using PyTorch library. 

\subsection{Metrics}
The performance of the classifiers were evaluated on the basis of their accuracy, precision, recall and F$_{1}$ score as given in (\ref{eq:Acc}), (\ref{eq:Precision}), (\ref{eq:Recall}), and (\ref{eq:f1}): 

\begin{equation}
Accuracy=\frac{tp+tn}{tp+fp+tn+fn}
\label{eq:Acc}
\end{equation}
\begin{equation}
Precision=tp/(tp+fn)
\label{eq:Precision}
\end{equation}
\begin{equation}
Recall=tp/(tp+fp)
\label{eq:Recall}
\end{equation}
\begin{equation}
F_1=\frac{2*precision*recall}{precision+recall}
\label{eq:f1}
\end{equation}
where true positives, true negatives, false positives and false negatives are denoted as $tp$, $tn$, $fp$, and $fn$, respectively.

% \usepackage{multirow}
\begin{table}[b!]
\centering
\caption{Performance metrics for the networks with the individual scans (T1w, FA, and MD) and the network with three channel inputs (T1w+DTI).}
\label{table:results1}
% \usepackage{multirow}
\begin{tabular}{lcccc}
\\ \hline
& T1w & FA & MD &  T1w+DTI    \\ \hline

Accuracy          & 0.82                 & 0.77                & 0.80                 & \textbf{0.97}      \\ \hline
\multicolumn{5}{l}{Precision}                                                                                                 \\
NC                & 0.78                 & 0.68                & 0.73                & \textbf{0.95}      \\
MCI               & 0.93                 & 0.94                & 0.94                & \textbf{0.98}    \\
AD                & 0.81                 & 0.74                & 0.77                & \textbf{0.98}     \\ \hline
\multicolumn{5}{l}{Recall}                                                                                                    \\
NC                & 0.95                 & 0.84                & 0.89                & \textbf{0.99}      \\
MCI               & 0.85                 & 0.92                & 0.89                & \textbf{0.99}      \\
AD                & 0.67                 & 0.57                & 0.62                & \textbf{0.93}       \\ \hline
\multicolumn{5}{l}{F1-score}                                                                                                  \\
NC                & 0.85                 & 0.75                & 0.80                 & \textbf{0.96}    \\
MCI               & 0.88                 & 0.93                & 0.91                & \textbf{0.98}      \\
AD                & 0.72                 & 0.64                & 0.68                & \textbf{0.95}     \\ \hline
\end{tabular}
\end{table}

\begin{table}[t!]
\centering
\caption{Results of the input agnostic network with T1w, FA, MD and three modality inputs.}
\label{table:results2}
% \usepackage{multirow}
\begin{tabular}{lcccc}
\\ \hline
& T1w  & FA   & MD   & T1w+DTI \\ \hline
Accuracy   & 0.86 & 0.69 & 0.78 & \textbf{0.96}    \\ \hline
\multicolumn{5}{l}{Precision}                                                                                                 \\
NC                &0.82 & 0.81 & 0.76 & \textbf{0.94}    \\
MCI              & 0.91 &\textbf{0.99} & 0.83 & \textbf{0.99}    \\
AD                &  0.86 & 0.62 & 0.92 & \textbf{0.96}    \\ \hline
\multicolumn{5}{l}{Recall}                                                                                                    \\
NC                & \textbf{0.96} & 0.68 & 0.94 & \textbf{0.96}    \\
MCI               &  0.89 & 0.46 & 0.87 & \textbf{0.97}    \\
AD                &  0.72 & 0.92 & 0.53 & \textbf{0.95}    \\ \hline
\multicolumn{5}{l}{F1-score}                                                                                                  \\
NC             & 0.88 & 0.69 & 0.83 & \textbf{0.95}    \\
MCI                 & 0.90  & 0.62 & 0.84 & \textbf{0.98}    \\
AD                  & 0.77 & 0.72 & 0.60  & \textbf{0.95}  \\ \hline
\end{tabular}
\end{table}

\section{Results and Discussion}
\label{resultsanddiscussion}
The results in Table~\ref{table:results1} show average values over five folds obtained from testing the networks with the individual scans (T1w, FA and MD) and the network with three channel inputs (T1w+DTI). We note that the fusion of T1W and DTI scans provide the highest accuracy (0.97). The accuracy of the network for the T1w images is higher than the networks for the FA and MD images. The results of the input agnostic network are provided in the Table~\ref{table:results2}. Similar to the individual networks, inputs consisting of both modalities perform highest while T1w images achieve the highest accuracy among the input types based on single image type.

When we compare the individual networks in Table~\ref{table:results1} with the input agnostic network in Table~\ref{table:results2}, we note that both approaches yield similar results. Specifically, the input agnostic network has better performance for T1w modality while the individual networks perform better for the other cases. These results highlight the potential of input agnostic models for AD classification. Presumably, additional modalities such as functional MRI and PET can also be integrated to the input agnostic network architecture and yield higher accuracy. 

In accordance with other studies on multi-modal AD diagnosis~\cite{aderghal2020improving,lee2019predicting,kang2020identifying}, using both modalities (T1w and DTI) resulted in better performance compared to using individual images (T1w, FA and MD). Our accuracy for using two modalities for both of our architectures is higher than the results presented in \cite{aderghal2020improving} (66.49\% accuracy) which also used both T1w and DTI modalities. This might be due to the simpler LeNet network architecture used in \cite{aderghal2020improving}. It might also be due to the datasets (OASIS-3 in our work versus ADNI in \cite{aderghal2020improving}). Only hippocampal region was used in \cite{aderghal2020improving} and this hints that features extracted from the whole brain might improve the classification accuracy.

Kang et al.~\cite{kang2020identifying} also used also transfer learning (from VGG16 network) to extract the features from T1w, FA and MD images and support vector machines (SVM) to discriminate MCI patients from NC using the ADNI dataset. Our results for three classes have a higher accuracy compared to their two class problem (94.2\% accuracy). This might be again related to the used datasets (OASIS-3 in our work versus ADNI in \cite{kang2020identifying}). This also suggests that extending the network with a fully connected layer for classification might be better than using another classifier (SVM in this case).

%\cite{marzban2020alzheimer} also developed 3-way (NC/MCI/AD) CNN classifier based on FA, MD and gray matter region of T1w images. However, their evaluation was done on binary discrimination, which hurdles the comparison with our three-way classifier. Besides,one of our objectives was to enable the diagnosis when both DTI and T1w images are not available, which is unlike to previous studies mentioned above and its performance is compatible with combination models. 

% \usepackage{multirow}
% \usepackage{multirow}

\section{Conclusions}
\label{conclusions}
In this work, we presented two deep learning architectures for the multimodal classification of three AD stages (NC, MCI, and AD) trained with data from the OASIS-3 dataset. Both our models outperform the ones in the literature trained using the ADNI dataset. Our input agnostic architecture enables the input of either T1W or DTI or both modalities into the same model. Our future work entails integrating other imaging modalities such as PET and functional MRI to the input agnostic model. 

\addtolength{\textheight}{-1cm}   % This command serves to balance the column lengths

%%%%%%%%%%%%%%%%%%%%%%%%%%%%%%%%%%%%%%%%%%%%%%%%%%%%%%%%%%%%%%%%%%%%%%%%%%%%%%%%

%%%%%%%%%%%%%%%%%%%%%%%%%%%%%%%%%%%%%%%%%%%%%%%%%%%%%%%%%%%%%%%%%%%%%%%%%%%%%%%%

%%%%%%%%%%%%%%%%%%%%%%%%%%%%%%%%%%%%%%%%%%%%%%%%%%%%%%%%%%%%%%%%%%%%%%%%%%%%%%%%
%\section*{APPENDIX}
%\section*{ACKNOWLEDGMENT}
%\begin{thebibliography}{99}
\bibliographystyle{IEEEtran}
\bibliography{main}
%\end{thebibliography}
\end{document}